%% file: aa1945.tex
\documentclass{aa}

\usepackage{graphicx}

\begin{document}

\newcommand{\Zsolar}{\mbox{${\; {\rm Z_{\sun}}}$}}
\newcommand{\ha}{\hbox{H$\alpha$}}
\newcommand{\oii}{\hbox{[O\,{\sc ii}]}}
\newcommand{\nii}{\hbox{[N\,{\sc ii}]}}
\newcommand{\sii}{\hbox{[S\,{\sc ii}]}}
\newcommand{\oiii}{\hbox{[O\,{\sc iii}]}}
\newcommand{\hb}{\hbox{H$\beta$}}
\newcommand{\hg}{\hbox{H$\gamma$}}
\newcommand{\hd}{\hbox{H$\delta$}}
\newcommand{\hi}{\hbox{H\,{\sc i}}}
\newcommand{\hii}{\hbox{H\,{\sc ii}}}
\newcommand{\heii}{\hbox{He\,{\sc ii}}}
\newcommand{\etal}{\hbox{et\thinspace al.\ }}
\newcommand{\oiiha}{\hbox{[O\,{\sc ii}]/H$\alpha$}}
\newcommand{\fha}{\hbox{$F_{{\rm H}\alpha}$}}
\newcommand{\fhb}{\hbox{$F_{{\rm H}\beta}$}}
\newcommand{\foii}{\hbox{$F_{\rm [O\,{\sc II}]}$}}
\newcommand{\micron}{\hbox{$\mu$m}}

\newcommand{\zoh}{\hbox{$12\,+\,{\rm log(O/H)}$}}

\newcommand{\teoii} {\hbox{$T_e{\rm (O\,{\sc II})}$}}
\newcommand{\tenii} {\hbox{$T_e{\rm (N\,{\sc II})}$}}
\newcommand{\teoiii}{\hbox{$T_e{\rm (O\,{\sc III})}$}}

\titlerunning{Oxygen abundances of BCGs}
\authorrunning{F. Shi \etal}

\title{Spectroscopic study of blue compact galaxies}
\subtitle{V. oxygen abundance and the metallicity-luminosity relation}

\author{F. Shi\inst{1}, 
        X. Kong\inst{1,2}, 
        C. Li\inst{1}, 
        F. Z. Cheng\inst{1}}
\mail{sfemail@mail.ustc.edu.cn;xkong@ustc.edu.cn}

\institute{
Center for Astrophysics, University of Science and Technology
of China, 230026, P. R. China
\and
National Astronomical Observatory, 2-21-1 Osawa, Mitaka, Tokyo 
181-8588, Japan
}
\date{Received; accepted}

\abstract{
This is the fifth paper in a series studying the stellar components, 
star formation histories, star formation rates and metallicities of 
a blue compact galaxy (BCG) sample. Based on our high-quality 
ground-based spectroscopic observations, we have determined the 
electron temperatures, electron densities, nitrogen abundances and 
oxygen abundances for 72 star-forming BCGs in our sample, using 
different oxygen abundance indicators.  
The oxygen abundance covers the range $7.15 < \zoh < 9.0$, and 
nitrogen is found to be mostly a product of secondary nucleosynthesis 
for $\zoh >8.2$ and apparently a product of primary nucleosynthesis for 
$\zoh < 8.2$.  
To assess the possible systematic differences among different oxygen 
abundance indicators, we have compared oxygen abundances of BCGs obtained with 
the $T_e$ method, $R_{23}$ method, $P$ method, $N2$ method and $O3N2$ 
method. The oxygen abundances derived from the $T_e$ method are 
systematically lower by 0.1--0.25 dex than those derived from the 
strong line empirical abundance indicators, consistent with previous 
studies based on \hii\ region samples. We confirm the existence of 
the metallicity-luminosity relation in BCGs over a large range of 
abundances and luminosities. Our sample of galaxies shows that the slope 
of the metallicity-luminosity relation for the luminous galaxies 
($\sim -0.05$) is slightly shallower than that for the dwarf 
galaxies ($\sim-0.17$). 
An offset was found in the metallicity-luminosity relation of the 
local galaxies and that of the intermediate redshift galaxies. It 
shows that the metallicity-luminosity relation for the emission line 
galaxies at high redshift is displaced to lower abundances, higher 
luminosities, or both.

}

\maketitle

\section{Introduction}\label{intro}

Spectra of BCGs are dominated by the emission of young, hot  star 
clusters that ionize their environment, and are characterized by 
their blue color, compact appearance, high gas content, strong 
nebular emission lines and low chemical abundances (Kunth \& 
{\" O}stlin 2000, Stasi{\' n}ska \etal 2001).  
Recent analyses of such objects, 
comparing the observed nebular emission lines, colors and stellar 
features with population synthesis models, found that most have 
experienced a recent, quasi-instantaneous burst of star formation 
(Mas-Hesse \& Kunth 1999; Noeske \etal 2000; Kong 2004). 
In addition, there is ample evidence for  older stellar populations 
in BCGs (Thuan 1983; Papaderos \etal 1996; Kong \etal 2003; Noeske 
\etal 2003).

Metallicity is a key parameter that controls many aspects in the 
formation and evolution of galaxies. The metallicity of BCGs is a 
parameter of recognized importance when trying to characterize 
their evolutionary status and link them to other objects showing 
overlapping properties, like dwarf irregular, low surface 
brightness galaxies, or high redshift compact galaxies (Hunter \& 
Hoffman1999; Contini \etal 2002; Izotov \etal 2004). Metal content 
is also at the base of global relations like those existing or 
searched for with the luminosity and the gas mass fraction 
(P{\'e}rez-Montero \& D{\'\i}az 2003; Kennicutt \etal 2003). 
Undergoing intense bursts of star formation, strong and 
narrow emission line spectra, relative small dust extinction, low 
metal environments and different star formation histories make 
BCGs appropriate laboratories to study the metallicity of galaxies 
(Kunth \& {\" O}stlin 2000).  We have undertaken an extensive 
long-slit spectral observation of BCGs. In the papers of this series, 
we have studied the stellar populations, star formation histories 
and star formation rates of BCGs. In the present paper, we determined 
the oxygen abundance for 72 star-forming BCGs in our sample, based 
on our high quality ground-based spectroscopic observations.

The determination of oxygen abundance is a critical stage prior to 
deriving the value for the metallicity in galaxies and the abundances 
for several other elements. The preferred 
method for determining the oxygen abundance in galaxies using \hii\ 
regions is through electron temperature-sensitive lines (the 
so-called $T_e$ method), such as the \oiii$\lambda$4363 line 
(Kennicutt \etal 2003).  However, for oxygen-rich galaxies, the 
oxygen line \oiii$\lambda$4363 is weak and difficult to detect.  
Alternative abundance determinations consist of empirical 
calibrations of the strong emission lines which are easily observable, 
such as the $R_{23}$ method, $P$ method, $N2$ method 
and $O3N2$ method (Pagel \etal 1979; Kobulnicky, Kennicutt, \& 
Pizagno 1999; Pilyugin \etal 2001; Charlot \& Longhetti 2001; 
Denicol{\'o} \etal 2002; Pettini \& Pagel 2003; Tremonti \etal 2004).  
These methods are based on the direct measurements of the electronic 
temperature of low metallicity galaxies and on theoretical models 
for high metallicity galaxies, without any direct electron 
temperature measurement. Based on our homogeneous BCG optical 
spectral sample, we determined the oxygen abundance of BCGs by both 
the $T_e$ method and those empirical methods. The results can be used 
to test the consistency of these different oxygen abundance 
indicators and to understand the physical origins of any systematic 
differences.

Interest in the relationship between luminosity (mass) and 
metallicity dates back several decades, begining with the seminal work 
of Lequeux \etal (1979).  A correlation between the metallicity and 
the blue luminosity for irregulars, spirals and ellipticals was 
demonstrated by various authors (Garnett \& Shields 1987; Skillman 
\etal 1989; Zaritsky, Kennicutt, \& Huchra 1994; Melbourne \& Salzer 
2002; Lamareille \etal 2004; Tremonti \etal 2004),  over $\sim$ 10 
magnitudes in luminosity and 2~dex in metallicity.  However, 
some recent  studies do not support these results for all types of 
galaxies.  In a sample 
of low surface brightness galaxies, McGaugh (1994) saw no 
relationship between $M_B$ and O/H. In a careful reanalysis of data 
using only the abundances determined from \oiii$\lambda4363$, 
Hidalgo-G\'{a}mez and Olofsson (1998) found no relationship between 
$M_B$ and O/H of irregular galaxies. Hunter \& Hoffman(1999) found 
that the relationship between $M_B$ and O/H for Im, Sm and blue 
compact dwarf galaxies has a very large scatter. 
The question is whether the metallicity-luminosity relation for 
dwarf galaxies exists in a similar manner as for massive galaxies. 
Using a sample of 519 star-forming emission-line galaxies from the 
KPNO International Spectroscopic Survey, Melbourne \& Salzer (2002) 
found that the slope of the metallicity-luminosity relation for 
luminous galaxies is steeper than that for dwarf galaxies. 
Using 1000 individual spectra of \hii\ regions in 54 late-type 
galaxies, however, Pilyugin \etal (2004) found that the slope of 
the metallicity-luminosity relationship for spirals ($M_B =-18 \sim 
-22$) is slightly shallower than the one for irregular galaxies 
($M_B = -12 \sim -18$). 
Using 72 star-forming BCGs (M$_B = -22 \sim -13$ mag), we will 
investigate the slope of the metallicity-luminosity relationship 
of BCGs.

We begin with a brief description of the spectroscopic observation 
and data reduction in Section 2.  We outline our method for 
measuring electron density and temperature in Section 3. We 
determine the oxygen abundance of BCGs by different methods in 
Section 4.  In Section 5 we compare our results with previous 
studies, and analyze the oxygen discrepancy between different 
methods, the luminosity--metallicity relation and the N/O -- O/H 
relation of BCGs. The conclusions are summarized in Section 6.

\section{Observations and data reduction}\label{obs-data}

To study the stellar components, star formation histories, star 
formation rates and metallicities of BCGs, we have prepared an 
atlas of optical spectra of the central regions of 97 BCGs in the 
first paper of this series (Kong \& Cheng 2002a). The spectra were 
obtained at the 2.16 m telescope at the XingLong Station of the 
National Astronomical Observatory of China.  A 300 line mm$^{-1}$ 
grating was used to achieve coverage in the wavelength region from 
3580 to 7400 \AA\ with about 4.8\,\AA\ per pixel resolution. The 
emission line equivalent widths and fluxes for our BCG sample were 
provided in the second paper of this series (Kong \etal 2002b). The 
typical uncertainties of the measurements are less than 10\% for 
\ha6563, \hb4861, and \oii\, emission lines.  The fluxes were 
dereddened for Galactic extinction, using the extinction 
coefficients from Schlegel \etal (1998) and the empirical 
extinction law from Cardelli \etal (1989).

To derive equivalent widths for underlying stellar 
absorption lines and to correct the measured Balmer emission line 
fluxes for these absorptions, we have applied an empirical 
population synthesis method to our BCG spectra (Kong \etal 2003).  
Intrinsic reddenings (attenuation of interstellar dust) were 
determined using the two strongest Balmer lines, \ha/\hb\ (Galactic 
extinction and underlying stellar absorption were corrected), and 
the effective absorption curve $\tau_{\lambda}$ = 
$\tau_V(\lambda/5500{\rm\AA})^{-0.7}$, which was introduced by 
Charlot \& Fall (2000). Using these intrinsic reddening values, the 
internal extinction of each galaxy was corrected.
The resulting emission line fluxes, which are corrected for 
the underlying stellar absorption as well as Galactic and internal 
extinction, will be used 
to determine the physical conditions (electron density, temperature,
oxygen and nitrogen abundance) of the galaxies in the next sections.

\section{Electron density and temperature}\label{teandne}

To derive oxygen element abundances with the $T_e$ method, we adopted 
a two-zone photoionized \hii\ region model (see Sec. \ref{teoxy}). 

The most precise method of determining the abundances of galaxies 
requires the detection of \oiii$\lambda$4363, an emission line that 
is weak and often not detected in oxygen-rich galaxies. 
In our spectra, \oiii$\lambda$4363 was detected in the spectra for 
45 BCGs, but only 20 of them have equivalent widths for 
\oiii$\lambda$4363 larger than 2{\rm\AA}, and uncertainties of 
the \oiii$\lambda$4363 flux measurements less than 20\%. 
For these 20 BCGs, the electron temperature \teoiii\ of the  
ionized gas was calculated using the method outlined by Shaw \& 
Dufour (1995) from the line flux ratio 
\oiii$\lambda$4363/($\lambda$4959+$\lambda$5007). For other BCGs, 
where the temperature sensitive line \oiii$\lambda$4363 is 
undetectable or has a low signal-to-noise ratio, an 
empirical relation of $T_e$ and strong spectral lines has been 
adopted for the electron temperature determination (Pilyugin 2001).  
The temperature will be used for derivation of the O$^{+2}$ ionic 
abundances. 

To estimate the temperature in the low-temperature zone \teoii, the 
relation between \teoii\ and \teoiii\ from Garnett (1992) is 
utilized : 
\begin{equation} 
t_e{\rm (O\,{\sc II})} = 0.7 \times t_e{\rm (O\,{\sc III})} + 0.3, 
\end{equation} 
where $t_e$=$T_e/10^4$ K. The temperature \teoii\ is 
used to derive the O$^+$ ionic abundance. 

Because the velocity dispersions of galaxies are typically hundreds 
of km s$^{-1}$, the classical density diagnostic 
\oii$\lambda\lambda3726,3729$ lines are barely resolved, so they 
do not place significant constraints on the density.  
In this paper, the electron number density $n_e$ of BCGs was 
determined by using the \sii$\lambda$6716/$\lambda$6731 line ratio.
In several cases where the sulfur lines are too noisy to accurately 
determine the electron density, we assume a density of 100 cm$^{-3}$. 
This assumption almost does not affect our oxygen abundance 
determination, since the effect of temperature is much larger than 
that of electron density, which can be inferred from Eq. (4). In all 
cases, we assume that the density does not vary significantly from 
high-ionization zone to low-ionization zone.
Calculations of the electron number density are carried out using
 a five-level statistical equilibrium model in the IRAF NEBULAR 
package (de Robertis, Dufour, \& Hunt 1987; Shaw \& Dufour 1995), 
which makes use of the latest collision strengths and radiative 
transition probabilities. 

The measured electron temperatures \teoiii, and electron number 
densities are presented, respectively, in the second and third 
columns of Table 1. 
                          
\input{aa1945-t1.tex}

\section{Determinations of oxygen abundance}\label{oxymea}

\subsection{ $T_e$ method }\label{teoxy}

To derive the oxygen abundance of BCGs, it has been assumed that 
the oxygen 
lines originate in two regions: a high-ionization zone with 
the temperature \teoiii, where \oiii\ lines originate; and a 
low-ionization zone with the temperature \teoii, where \oii\ lines 
originate.  From the weakness of \heii$\lambda$4686, we know that 
a negligible fraction of the gas is in O$^{+3}$, so the oxygen 
abundance is simply the sum (O$^{+}$+O$^{++}$) by the expressions 
from Pagel \etal (1992):

\begin{equation}
\frac{\rm O}{\rm H} = \frac{O^+}{H^+} + \frac{O^{++}}{H^+},
\end{equation}
\begin{eqnarray}
12+ \log (O^{++}/H^+) = \log 
\frac{I(\oiii\lambda\lambda4959,5007)}
{I(\hb)} + \nonumber  \\ 
6.174 + \frac{1.251}{t_e{\rm (O\,{\sc III})}}  - 
0.55 \log t_e{\rm (O\,{\sc III})},
\end{eqnarray}
\begin{eqnarray}
12+ \log (O^{+}/H^+) = \log \frac{I(\oii\lambda3727)}
{I(\hb)} + 5.890 + \nonumber  \\
 \frac{1.676}{t_e{\rm (O\,{\sc II})}}  -
 0.40 \log t_e{\rm (O\,{\sc II})} + \log (1+1.35x)  ,
\end{eqnarray}
where $n_e$ is the electron density in cm$^{-3}$, 
and $x= 10^{-4} n_e t_e{\rm (O\,{\sc II})}^{-1/2}$.

For 70 out of 72 galaxies in our sample, we have measured the 
oxygen abundance from the $T_e$ method, and present them in 
the fourth column of Table 1. For the remaining two BCGs, III 
Zw 42 and I Zw 101, the  \oiii$\lambda5007$ line has a too low 
signal-to-noise ratio for reliable flux measurements, and the 
oxygen abundances cannot be determined by the $T_e$ method.  

\subsection{$R_{23}$ method}

The key for accurate determination of oxygen abundances in galaxies
by the $T_e$ method is a precise measurement of the weak auroral 
forbidden emission line \oiii$\lambda$4363. 
In star formation galaxies the temperature-sensitive 
\oiii$\lambda$4363\ line intensity correlates with the overall 
abundance, being relatively strong in very low metallicity systems 
and becoming undetectable even for moderately low metallicity 
galaxies (e.g. metallicity higher than 0.5 solar metallicity). As 
a result, for most of the star formation galaxies \oiii$\lambda$4363 
is unmeasurably weak (McGaugh 1991). 

To overcome this problem, Pagel \etal (1979) suggested the empirical 
abundance indicator
\begin{equation}
R_{23} = 
\frac{I(\oii\lambda3727) + 
I(\oiii\lambda\lambda4959,5007)}{I(\hb)}
\end{equation}
for high metallicity galaxies. 
The method has been subsequently refined and calibrated using both 
observational data and models (e.g. McGaugh 1991). Kobulnicky et al. 
(1999) showed that it can be as precise as 0.2 dex.
      
A major difficulty associated with this method is that the relation 
between oxygen abundance and $R_{23}$ is double valued, requiring 
some assumption or rough a priori knowledge of a galaxy's 
metallicity in order to locate it on the appropriate branch of the 
curve.  
In this work, the \nii$\lambda$6584/\ha\ line ratio will be used 
to break the degeneracy of the $R_{23}$ relation (Denicol\'o \etal 
2002). The division between the upper and the lower branch of the 
$R_{23}$ relation occurs around log$(\nii\lambda6583/\ha) \simeq 
-1.26$ (\zoh $\simeq 8.2$).
We use the most recent $R_{23}$ analytical calibrations given by
Kobulnicky et al. (1999) which are based on the models by McGaugh
(1991) to determine the oxygen abundances of BCGs in our sample. 
The computed oxygen abundances by the $R_{23}$ method are shown in 
the fifth column of Table 1.

\subsection{$P$ method}

New methods for abundance determinations using strong lines have 
been developed recently. These methods achieve a good approximation 
to the results obtained with the $T_e$ method.  

One of these new calibrations, the $P$ method, was proposed by 
Pilyugin (2000, 2001).  
By comparing  oxygen abundances in H II regions derived with the 
$T_e$ method and those derived with the $R_{23}$ method, the author 
found that the error in the oxygen abundance derived with the 
$R_{23}$ method involves two parts, a random error and a systematic 
error. The origin of this systematic error is the dependence of the 
oxygen emission lines on not only the oxygen abundance, but also 
on the other physical conditions (hardness of the ionizing radiation 
and a geometrical factor).

To determine accurate abundances in HII regions and galaxies, 
Pilyugin derived a new relation between the oxygen abundance and the 
value of the abundance index $R_{23}$, the excitation parameter $P$. 
The best fitting relations
\begin{equation}
12 + \log ({\rm O}/{\rm H})_P = 
\frac{R_{23}+54.2+59.45P+7.31P^2}{6.07+6.71P+0.37P^2+0.243R_{23}}
\end{equation}
can be adopted for oxygen abundance determinations in moderately 
high-metallicity HII regions and galaxies, \zoh $\ge 8.2$, with 
undetectable temperature-sensitive line ratios; and 
\begin{equation}
12 + \log ({\rm O}/{\rm H}) = 6.35 + 1.45 \log R_3 - 3.19 \log P
\end{equation}
can be adopted for oxygen abundance determinations in moderately 
low-metallicity HII regions and galaxies, with 
$R_3 =I(\oiii\lambda\lambda4959,5007) / I(\hb)$, and $P=R_3/R_{23}$.

The oxygen abundances of 59 BCGs have 
$\log [I(\nii)\lambda6583/I(\ha)] > -1.26$ (see in the next section), 
they seem to belong to the high metallicity branch calibration. The 
oxygen abundances of these 59 galaxies were calculated through 
Eq.~(6). 11 of 72 BCGs belong to the low metallicity branch, Eq.~(7) 
was used for their oxygen abundance calibration. III Zw 42 and I Zw 
101 have very weak \oiii$\lambda$5007 emission lines; we did not
determine their oxygen abundances. 
The oxygen abundances computed with the $P$ method are shown in the 
sixth column of Table 1.

\subsection{$N2$ method}

Following the earlier work by Storchi-Bergmann, Calzetti, \& 
Kinney (1994) and by Raimann \etal (2000), Denicol{\'o} \etal (2002) 
 focused attention on the 
$N2 \equiv \log [I(\nii\lambda6583) / I(\ha)]$ index. 
They collected a representative sample of spectroscopic measurements 
of star forming galaxies covering a wide range in metallicity (7.2 
$\le$ \zoh $\le$ 9.1) from the literature, and recalculated oxygen 
abundances in a self-consistent manner with a precision of $\sim$ 
0.2 dex. 

The $N2$ and the oxygen abundance are well correlated (linear 
correlation coefficient of 0.85) and a single slope is capable 
of describing the whole metallicity range, from the most metal-poor  
to the most metal-rich galaxies in the sample. Least squares fits 
to the data simultaneously minimizing the errors in both axes give 
\begin{equation}
\zoh = 9.12 + 0.73 \times N2  
\end{equation}

Because the $N2$ vs. metallicity relation is monotonic, and the $N2$ 
line ratio does not depend on reddening corrections or flux 
calibration, the $N2$ indicator was used to determine the oxygen 
abundance for our BCGs, and also to break the degeneracy of the 
$R_{23}$-(O/H) (in Sec. 4.2) and the $P$-(O/H) (in Sec. 4.3) relation. 
The measured oxygen abundances for BCGs by the $N2$ method are shown 
in the seventh column of Table 1.

\subsection{$O3N2$ method}

Alloin \etal (1979) was the first to introduce the quantity 
$O3N2 \equiv \log$ $\{[I(\oiii\lambda5007)/I(\hb)]$ / 
$[I(\nii\lambda6583)/I(\ha)]$\}\footnote{This definition is 
slightly different from the original one proposed by Alloin \etal 
(1979) who included both \oiii\ doublet lines in the numerator of 
the first ratio.}, but since then the $O3N2$ index has been 
comparatively neglected in nebular abundance studies (Pettini \& 
Pagel 2004). 

Pettini \& Pagel(2004) reconsidered the $O3N2$ vs.~metallicity 
relation, using 137 extragalactic \hii\ regions. The sample is 
similar to that of Denicol{\'o} \etal (2002). They found that at 
$O3N2 \le 1.9$, there appears to be a relatively tight, linear and 
steep relationship between $O3N2$ and $\log {\rm (O/H)}$. A least 
squares linear fit to the data in the range $-1 < O3N2 < 1.9$ 
yields the relation:
\begin{equation}
12 + \log {\rm (O/H)} = 8.73 - 0.32 \times O3N2.
\end{equation}
This relationship is valid only when $O3N2 \le 1.9$. 61 out of the 
72 BCGs in our sample satisfy this condition, and the oxygen 
abundances are estimated from the $O3N2$ indicator, and are shown 
in the eighth column of the Table 1.

\section{Discussions}\label{discu}

\subsection{Oxygen abundance uncertainties}

The error of the value of oxygen abundances derived with the 
$T_e$-based method and the strong line empirical methods 
involves two parts: a random error and a systematic error.

The main source of random errors in the oxygen abundance 
determination is the uncertainty in the intensities of the lines
used to derive the abundances.
Since we have corrected the effect of the underlying stellar 
absorption (by an empirical population synthesis method) and the 
dust extinction accurately, these effect can be neglected.
The typical uncertainties of the measurements are less than 10\% 
for most of the emission lines.  
Using the error of the line fluxes and standard formulae of 
propagation of errors, we estimate the corresponding random errors of the
oxygen abundances for our galaxies.  
The typical error in the oxygen abundances is 0.17 dex, 0.11 dex,
0.08 dex; 0.06 dex and 0.06 dex by the $T_e$ method, $R_{23}$ 
method, $P$ method, $N2$ method and $O3N2$ method, respectively.
The typical systematic error, using the scatter of the calibrations 
from the respective literature, in the oxygen abundances by the 
$T_e$ method is less than 0.1 dex, and is $\sim$ 0.2 dex by the 
strong line empirical methods. 

The other uncertainty of oxygen abundances from the $R_{23}$ method 
and the $P$ method is the double-value relation between oxygen 
abundance and $R_{23}$. 
To break this degeneracy, the $N2$=\nii$\lambda$6584/\ha\ line ratio 
was used to discriminate between the high-- and low-- metallicity
ranges of the $P$ and $R_{23}$ method calibrations. Given the 
scatter of the $N2$ vs. O/H relation, objects close to $N2=-1.26$ 
may be either in the high or low metallicity branch of the $P$ and 
$R_{23}$ calibrations.
When analyzing the galaxies in our sample, we found most of them 
have high or low metallicity, only 6 of them close to $N2=-1.26$.
Therefore, the double-value relation between oxygen abundance and 
$R_{23}$ does not affect the oxygen abundances from the $R_{23}$ and 
the $P$ method for the most galaxies in our sample.

\subsection{Comparison with previous studies}

\begin{figure}
\centering
\includegraphics[angle=-90,width=\columnwidth]{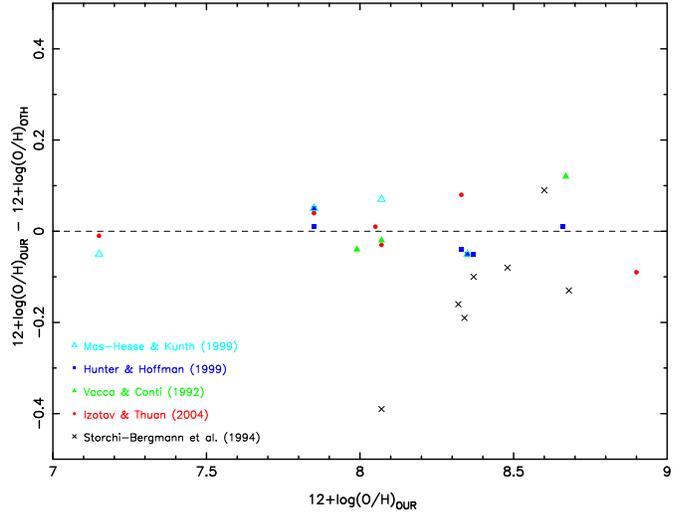}
\caption{
Comparison of the oxygen abundances in our work with those in 
previous studies. The oxygen abundances were derived from the 
$T_e$-method.  
Filled triangles, crosses, filled squares, open triangles, and 
filled circles represent galaxies from Vacca \& Conti (1992), 
Storchi-Bergmann, Calzetti, \& Kinney (1994), Hunter \& Hoffman 
(1999), Mas-Hesse \& Kunth (1999), and Izotov \& Thuan (2004), 
respectively. The dashed line is the equal-value line.
}
\label{zour-zoth}
\end{figure}

In the last section, we have determined the oxygen abundance of BCGs 
in our sample, using different oxygen abundance indicators. In this 
section, we will discuss the differences between the results from 
these oxygen abundance indicators and the absolute magnitude--oxygen 
abundance relationship. Before this, it will 
be useful to compare the oxygen abundance determined from the $T_e$ 
method in this work and in previous studies. 

For 20 out of 72 star-forming BCGs in our sample, $T_e$-based oxygen 
abundance have been estimated in previous studies. These oxygen 
abundances are shown in the last column of the Table 1. The 
differences between the oxygen abundance estimated in our work, 
\zoh$_{\rm OUR}$, and those in other studies, \zoh$_{\rm OTH}$, 
are shown in Figure \ref{zour-zoth} with different symbols. 
The oxygen abundance in our work is in good agreement with the value 
of Vacca \& Conti (1992), Hunter \& Hoffman (1999), Mas-Hesse \& Kunth 
(1999), and Izotov \& Thuan (2004), the differences are usually less 
than 0.1 dex.  
However, the discrepancy between the oxygen abundances in our work and 
those in Storchi-Bergmann, Calzetti \& Kinney (1994) is large.  
One possible reason for this discrepancy are the different observation 
aperture sizes, a relatively large aperture ($10\arcsec \times 20\arcsec$) 
having been used in Storchi-Bergmann, Calzetti \& Kinney (1994) and a 
small slit aperture (slit width $\sim 2 \arcsec$) in our work and in 
Vacca \& Conti (1992), Hunter \& Hoffman (1999), Mas-Hesse \& Kunth 
(1999) and Izotov \& Thuan (2004), if there exists a radial 
metallicity gradient or line flux/ionization gradient in BCGs.

\subsection{Oxygen abundance from different indicators}

\begin{figure*}
\centering
\includegraphics[angle=-90,width=0.9\textwidth]{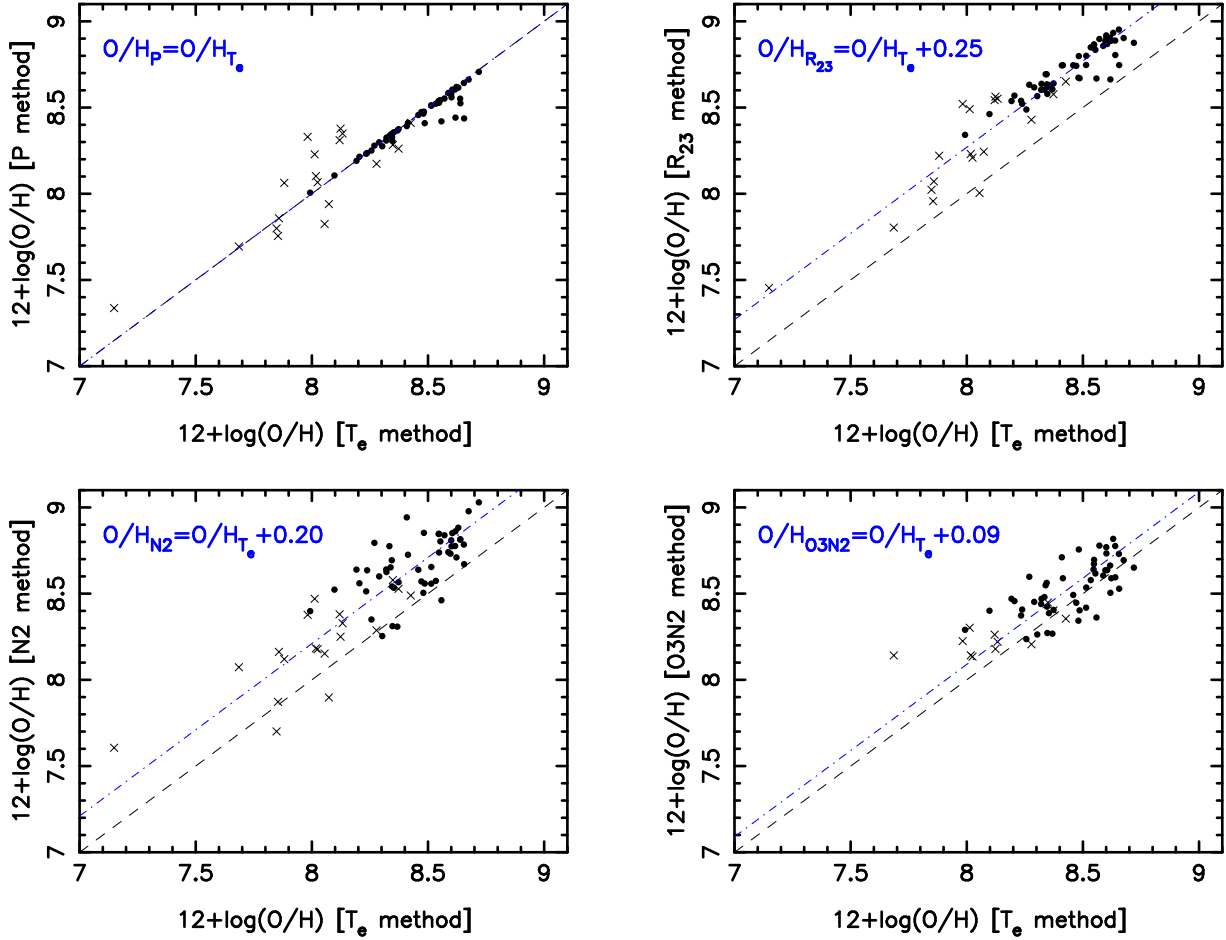}
\caption{
Comparison of the electron temperature based abundances with those 
derived from different strong line calibrations.  
Filled circles represent galaxies with temperatures determined from 
the strong spectral lines (Pilyugin 2001), while crosses are used 
for galaxies with temperatures determined from the temperature 
sensitive line \oiii$\lambda$4363.  
The four panels show the empirical abundances using the calibrations 
of the $P$ method (Pilyugin 2000; 2001), $R_{23}$ method (Kobulnicky 
\etal 1999), $N2$ method (Denicol{\'o} \etal 2003) and $O3N2$ method 
(Pettini \& Pagel 2004).  
The dashed line is the equal-value line, and the dot-dashed line is 
the best fitting linear relationship, where the fit was forced to 
have the same slope (equal to 1) as the equal-value line and was 
allowed to vary only by a constant, 
O/H$_{XXX}$ = O/H$_{\rm T_e}$ + OFFSET, O/H = \zoh, which can 
be found in the upper-left corner of each panel.
}
\label{metal-comp}
\end{figure*}

Abundance estimates based on the strong line empirical methods have 
largely supplanted the direct, $T_e$-based determinations for 
large-scale abundance surveys and cosmological lookback studies.  
With the large range in oxygen abundance in our BCG sample, it is 
instructive to study the difference of the oxygen abundance from the 
$T_e$-based method and from the empirical determination methods.

Figure \ref{metal-comp} shows the plot of the $T_e$-based oxygen 
abundance against the empirical method oxygen abundances for the blue 
compact galaxies in our sample.  
Galaxies with temperatures determined from the temperature sensitive 
line \oiii$\lambda$4363 are presented as crosses, and those with 
temperatures from the strong spectral lines (following Pilyugin \etal 
2001, see Section 3 of this paper) are presented as filled circles. 
Both abundances with temperatures determined from the 
\oiii$\lambda$4363 and from the strong spectral lines are regarded as 
the standard $T_e$--based abundance and used to study the offset 
between the P--, $R_{23}$--, $N2$--, $O3N2$--method abundances and 
the $T_e$--based abundances.

It shows that although with a large scatter, there is an agreement
between the oxygen abundance from the $T_e$ method and those from
the empirical methods for these low metallicity galaxies whose 
temperatures are determined from the \oiii$\lambda$4363 line. But for
high metallicity galaxies, there exists an offset between the oxygen
abundance from the empirical methods and the $T_e$--based method. 

Fig.\ref{metal-comp}a shows that there is a good agreement between 
the O/H$_{T_{e}}$ and the O/H$_{P}$ abundances for high metallicity 
galaxies.  Because the temperature of these high metallicity 
galaxies was determined by the $P$ method too (Pilyugin 2001), this 
agreement can be understood easily, and cannot be used to test the 
validity of the $P$ method.  
For the other indicators, although all the $T_e$-based galaxy 
abundances trace a locus which is roughly consistent in shape with 
the empirical calibrations, there is a pronounced offset in 
abundance in most cases, as has been pointed out previously 
(Stasi\'nska 2002; Kennicutt \etal 2003). 
In all cases, the empirical calibrations yield oxygen abundances 
that are systematically higher than the $T_e$-based abundances, by 
amounts ranging from 0.09 to 0.25 dex on average. The systematic 
offset may have significant consequences for the nebular abundance 
scale as a whole. If the $T_e$ abundances are correct, it implies 
that most studies of the galactic abundances (locally and at high 
redshift) based on empirical nebular line calibrations have 
over-estimated the true absolute oxygen abundances by factors of
1.2--1.8 for high metallicity galaxies, as was noted by Kennicutt
\etal (2003).

The discrepancies shown in Figure~\ref{metal-comp} can be traced 
to two main origins, an insufficient number of calibrating \hii\ 
regions with accurate $T_e$-based abundances in the earliest 
calibrations of the empirical methods and a systematic offset 
between the nebular electron temperatures in the calibrating 
photoionization models and the observed forbidden-line temperatures 
for a given strong line spectrum (Kennicutt \etal 2003). Detailed 
discussions of the reasons of these discrepancies can be found in 
Kennicutt, Bresolin \& Garnett (2003). 

Currently, we cannot be certain whether the discrepancies in 
abundance scales are due to the biases in the $T_e$-based results, 
or the problems in the theoretical models that are used to calibrate 
most of the strong line empirical abundance indicators. High-quality 
far-infrared measurements of a sample of extragalactic HII regions, 
including some of the principal fine-structure cooling lines, may 
help to resolve these inconsistencies.

\subsection{The luminosity-metallicity relationship}

\begin{figure*}
\centering
\includegraphics[angle=-90,width=0.9\textwidth]{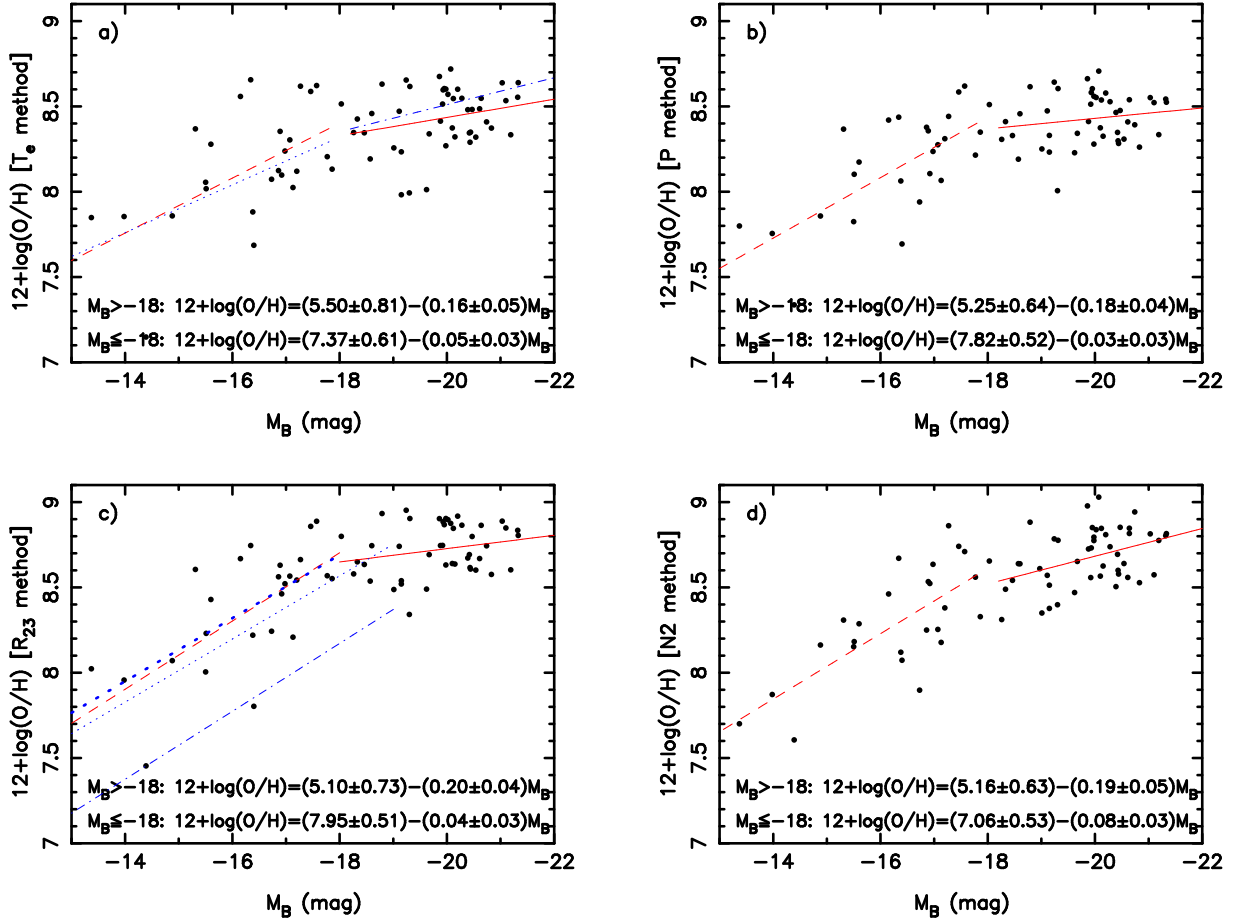}
\caption{
Luminosity--metallicity relations for BCGs in our sample, the oxygen 
abundances were derived from methods in Sec.~4. The solid line is a 
linear least squares fit to the luminous galaxies ($M_B \leq -18$), 
the dashed line is a linear least squares fit to the dwarf galaxies 
($M_B > -18$, blue compact dwarf galaxies). 
The dotted and dash--dotted lines in a) show the relation from 
Pilyugin \etal (2004a) for the irregular and spiral galaxies. 
The dotted line in c) shows the relation for about 50,000 SDSS 
star--forming galaxies from Tremonti \etal (2004), the thick-dotted 
line in c) shows the best fitting line for our dwarf galaxies with 
the same slope as Tremonti \etal (2004), but with different offset,  
and the dash--dotted lines in c) shows the relation of the
 intermediate redshift galaxies from Maier \etal (2004).
}
\label{metal-MB}
\end{figure*}

Using different oxygen abundance indicators, the oxygen abundances 
of BCGs have been determined in the last section. Combining oxygen 
abundance and absolute magnitude of galaxies, we can study the 
luminosity--metallicity relationship of BCGs.

Figure \ref{metal-MB} shows the absolute magnitude $M_B$ (Kong \& 
Cheng 2002a) vs. oxygen abundance relation for BCGs in our sample.  
The general trend, widely discussed in the literature and confirmed 
here, is an increase of metallicity with luminosity over a large 
magnitude range, from $M_B=-13$ to $-22$.  
We also found that there is not a tight correlation between the two 
quantities. The origin of this scatter may be due to the differences 
in the star formation history, the evolutionary status of the 
current starburst, different initial mass function, starburst-driven 
outflows, winds, gas infall or due to the errors in metallicity 
determination. In addition, the relationship for the luminous 
galaxies ($M_B \leq -18$) is shallower than that of the dwarf 
galaxies ($M_B > -18$).  

To check any dependency of the luminosity--metallicity relation on 
the choice of the O/H calibration, the oxygen abundances determined 
by the $T_e$, $P$, $R_{23}$, and $N2$ methods were plotted in the 
different panels of Figure~\ref{metal-MB} (the abundance from the 
$O3N2$ method is not used, since this method does not work well for 
low metallicity dwarf galaxies). 
In each case, we performed a linear regression on the 
luminosity--metallicity relation for the luminous galaxies 
($M_B \leq -18$) and the dwarf galaxies ($M_B > -18$).
We does not consider the errors of $M_B$ and the
 fits were weighted by the errors of the O/H data.
The solid line is the luminosity--metallicity relation for the 
luminous galaxies, 
and the dashed line is the best fitting line for the dwarf galaxies.  
We found that the slopes of the different luminosity-metallicity 
relations are similar; the choice of oxygen indicator has a small 
effect on the slope of the luminosity--metallicity relation. However, the
different methods for oxygen abundances determined can introduce  
 zeropoint differences in the luminosity--metallicity 
relations.

It is instructive to compare our determination of the 
luminosity--metallicity relation with other published determinations, 
both to validate the consistency of our measurements and to compare 
the luminosity--metallicity relation in the different types of  galaxies. 
In Fig.~\ref{metal-MB}a, we plot the luminosity--metallicity relation 
for the spiral galaxies (dot-dashed line) 
\begin{eqnarray}
12 + \log {\rm (O/H)}_P&=&(6.93 \pm 0.37)-(0.079 \pm 0.018) \times M_B  \nonumber \\
(M_B < -18), 
\end{eqnarray}
and for the local irregulars (dotted line)
\begin{eqnarray}
12 + \log {\rm (O/H)}_P&=&(5.80 \pm 0.17)-(0.139 \pm 0.011) \times M_B  \nonumber \\
(M_B > -18), 
\end{eqnarray}
derived by Pilyugin \etal (2004a).
Inspection of Fig.~\ref{metal-MB}a shows that the 
luminosity--metallicity relationship for BCGs in our sample agrees, 
within the uncertainties, with that from Pilyugin \etal (2004a).
We confirm the existence of the difference of the slopes for the 
luminosity--metallicity relationships between luminous and dwarf 
galaxies.

Using a sample of 53,400 star-forming galaxies at \mbox{$z\sim0.1$} 
in the Sloan Digital Sky Survey, Tremonti \etal (2004) derived a 
luminosity--metallicity relation 
\begin{equation}
12 + \log {\rm (O/H)}=(5.238 \pm 0.018)-(0.185 \pm 0.001) \times M_B 
\end{equation}
for nearby galaxies($M_B > -21$ mag) using the new techniques that 
make use of the approach outlined by Charlot \etal (in 
preparation). This consists of estimating metallicity statistically, 
based on simultaneous fits of all the most prominent emission lines 
(\oii, \ha, \oiii, \hb, \nii, \sii) with a model designed for the 
interpretation of integrated galaxy spectra (Charlot \& Longhetti 
2001). This relation was plotted in Fig. \ref{metal-MB}c as a dotted 
line. Fig.~\ref{metal-MB}c shows that the slope of Tremonti \etal (2004)  
is almost same as that of ours, but there is a small offset between 
the curves. The small offset can be derived by 
fitting lines with a slope forced to the value by Tremonti \etal (2004). 
As a result, our new luminosity--metallicity relation becomes 
\begin{eqnarray}
12 + \log {\rm (O/H)}_{R_{23}}& =
& (5.36 \pm 0.72) -0.185 \times M_B \nonumber \\
(M_B > -18),
\end{eqnarray}
and is presented in Fig.~\ref{metal-MB}c as a thick-dotted line. 
A zeropoint offset, $\Delta$log(O/H) $\sim 0.12$, between Tremonti 
\etal (2004) and ours was found. The method differences should not 
induce this offset because the $R_{23}$ method and the method used 
by Tremonti \etal (2004) are both based on the similar strong lines.
The different redshift range in our sample ($z\sim0.0$) and that in 
Tremonti \etal (2004) ($z\sim0.1$) may cause this offset. 
However, the significance is not very high (see the 
errors in Eq. 13). To further assess this possible trend with 
redshift, we also plotted the luminosity--metallicity relation 
\begin{eqnarray}
12 + \log {\rm (O/H)}_{R_{23}} &=& (4.59 \pm 0.91)-(0.20 \pm 0.046)\times M_B \nonumber \\
(M_B > -19),
\end{eqnarray}
for emission line galaxies at intermediate redshifts ($z \simeq 
0.4$) in Fig.~\ref{metal-MB}c as dashed-dotted line (Maier \etal 2004). 
A larger zeropoint offset, $\Delta$log(O/H) $\sim 0.51$, was found for 
these intermediate redshift galaxies. 

Comparison between the luminosity--metallicity relationships in the 
different redshift ranges shows that the luminosity--metallicity 
relation at high redshift is displaced to lower abundance and higher 
luminosities compared to today. One explanation could be that 
intermediate-redshift galaxies are slightly less advanced in their 
evolution and, as a consequence, are slightly more metal-deficient 
than local galaxies of the same luminosity. As an alternative, 
intermediate-redshift galaxies may have just undergone a powerful 
starburst which temporarily increases their blue luminosity 
(Pilyugin \etal 2004b).  
                                                                          
\subsection{The N/O versus O/H relationship}

The origin of nitrogen has been a subject of debate for some years.  
Nitrogen is thought to be synthesized in the CNO process during 
 hydrogen burning. However the stars responsible remain uncertain 
(Izotov \& Thuan 1999, for short as IT99; Kunth \& {\"O}stlin 2000;
Contini \etal 2002).  
In the case of {\it secondary} synthesis, oxygen and carbon have 
been produced in the previous generations of stars, and the nitrogen, 
produced in the present generation of stars, should be proportional 
to their initial heavy element abundance. Secondary nitrogen 
production is expected in stars of all masses (see IT99). 
In the case of {\it primary} nitrogen synthesis, on the other hand, 
oxygen and carbon are produced in the same stars prior to the CNO 
cycle rather than in previous generations, and nitrogen production 
should be independent of the initial heavy element abundance. 
Primary nitrogen production is thought to occur mainly in 
intermediate-mass stars, yet important contributions may also come 
from high mass stars (see for example Weaver \& Woosley 1995, IT99, 
Izotov \etal 2004, and references therein).
Therefore, the N/O abundance ratio as a function of the O/H ratio 
is a key relation for understanding the origin of nitrogen of 
galaxies.

\begin{figure}
\centering
\includegraphics[angle=-90,width=0.9\columnwidth]{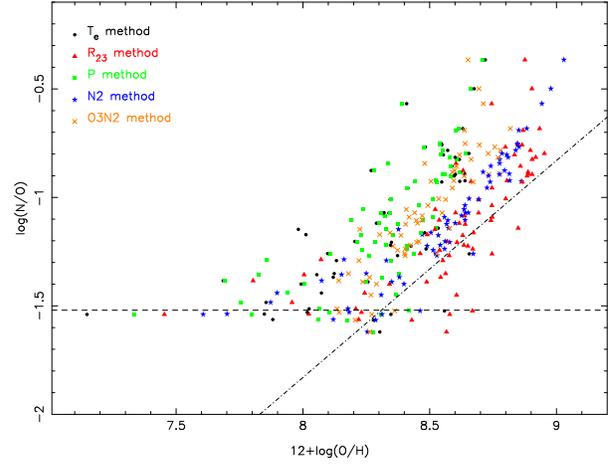}
\caption{
Logarithm of the number ratio of nitrogen to oxygen abundances 
plotted against \zoh.  Different symbols represent oxygen abundances 
from different methods, as shown in the upper-left corner of the 
figure and in Sec. 4. The dashed line ({\it primary} origin) and the 
dot-dashed line ({\it secondary} origin) represent the mean relation 
for galactic dwarf stars (Tomkin \& Lambert 1984). 
The points are not corrected for the mean O/H offsets derived for the 
different O/H calibrators in Section 5.3}
\label{rno}
\end{figure}

To understand the origin of nitrogen in BCGs, we plot the 
distributions of \zoh\ and log(N/O) abundance ratios for our sample 
galaxies in Figure~\ref{rno}. The N/O abundance ratios in BCGs 
were determined from the expression (Pagel \etal 1992):
\begin{eqnarray}
\log (N/O) = \log (N^+/O^+) 
= \log \frac{I(\nii\lambda\lambda6548,6584)}
{I(\oii\lambda3727)} \nonumber  \\
 +0.31-\frac{0.726}{t_e(\nii)}-0.02\log t_e(\nii)-
\log\frac{1+1.35x}{1+0.12x},
\end{eqnarray}
where 
$t_e(\nii) = t_e(\oii)$, and $x= 10^{-4} n_e t_e(\nii)^{-1/2}$.
Fitting lines, based on the solar neighborhood dwarf stars, from 
Tomkin \& Lambert (1984) for a primary (dashed line), and a 
secondary (dot-dashed line) production of nitrogen are shown in 
Figure~\ref{rno} too. 

To show the dependence of the relation between N/O and O/H on the 
methods for determining oxygen abundance, the oxygen abundances 
determined by both the $T_e$-based method and the strong line 
empirical abundance methods are shown in Figure~\ref{rno}. It 
should be noted 
that O/H from methods based on \nii\ line (such as the $N2$ and the 
$O3N2$ methods) are suspicious because these O/H are empirical and 
nitrogen-dependent.
Therefore, we draw conclusions from the $T_e$, P, $R_{23}$ method,  
which should be independent of nitrogen. 

In our sample, only one galaxy, I Zw 18, has a very low metallicity, 
\zoh $< 7.6$, and the log(N/O) of I Zw 18 is about $-1.5$, which is 
consistent with that in IT99 and references therein, and the 
fitting lines from Tomkin \& Lambert (1984) for a primary 
production of nitrogen.  
For galaxies with moderately low metallicity, $7.6 <$ \zoh $< 8.2$, 
 Fig.~\ref{rno} shows that the N/O ratio begins to increase with 
the oxygen abundance above $\sim -1.50$ along with the scatter. 
For galaxies with \zoh $ < 7.6$, IT99 interpreted these results
as primary nitrogen production by massive stars only. In this
scenario, galaxies in that O/H range are young, so that primary 
nitrogen from intermediate mass stars has not been released yet. 
Only in slightly older galaxies, do intermediate-mass stars start 
to contribute secondary nitrogen, leading to the higher mean and 
scatter of N/O ratios seen for $7.6 <$ \zoh $< 8.2$ . 
For an alternative
interpretation, see e.g. Izotov \etal (2004) and references therein.
For the galaxies with high metallicity, \zoh $> 8.2$, the N/O ratio 
increases with the oxygen abundance more rapidly, indicating that, in 
this metallicity regime, nitrogen is primarily a {\it secondary} 
element, and the contribution from the primary production is not 
important. The dispersion of N/O scatter is roughly about $\pm0.3$ 
dex. This scatter looks similar to that  obtained by IT99. 

\section{Summary and conclusions}\label{conclu} 

We determined the electron temperatures, electron 
densities, nitrogen abundance and oxygen abundance for 72 
star-forming blue compact galaxies in our sample, using different 
oxygen abundance indicators. The discrepancies of the oxygen 
abundances from different indicators, the relations between the 
oxygen abundance and luminosity and the N/O vs. O/H were 
investigated. We obtained the following results.

\begin{enumerate}
\item
The oxygen abundance in our BCGs ranges from \zoh = 7.15 to 9.0. The 
oxygen abundance derived from the $T_e$-based method in our work is 
in good agreement with the values of previous studies; the 
differences between them are usually less than 0.1 dex where observed 
aperture sizes are comparable.

\item
To study how significantly the oxygen abundance depends on the 
derivation method, the oxygen abundance of BCGs was measured 
with the $T_e$-based method and the strong line empirical abundance 
methods (such as the $R_{23}$, $P$, $N2$ and $O3N2$). We found that 
the empirical calibrations yield oxygen abundances that are 
systematically higher than the $T_e$-based abundances, by amounts 
ranging from 0.09 to 0.25 dex on average.

\item
We have confirmed the existence of the metallicity-luminosity 
relation in BCGs. 
It shows that the slope of the metallicity-luminosity relation for 
the luminous galaxies ($M_B < -18$) is shallower than that for the 
dwarf galaxies($M_B > -18$), and the metallicity-luminosity relation 
for emission line galaxies at high redshift is displaced to lower 
abundances, higher luminosities, or both, compared to BCGs in the local 
Universe.

\item
The relation between the N/O and the O/H abundance ratios from our
 data confirms the results of previous work (IT99). In the high
 metallicity regime, \zoh $ > 8.2$, nitrogen is primarily a
secondary element. For galaxies with lower metallicities, nitrogen
originates mainly from primary production in intermediate and high
 mass stars.
\end{enumerate}

\begin{acknowledgements}

We thank L.S. Pilyugin, A. C. Phillips for their helpful suggestions. 
The referee, K. G. Noeske, is thanked for the constructive report,
which helped improve the paper.
This work is based on observations made with the 2.16m telescope 
of the National Astronomical Observatory of China. XK acknowledges 
support provided by the Japan Society for the Promotion of Science 
(JSPS). 
\end{acknowledgements}

\end{document}

%% file: aa1945-t1.tex
\setcounter{table}{0}
\begin{table*}[]
\caption{Electron temperatures, electron densities, oxygen abundances 
and nitrogen to oxygen abundances of BCGs.}
\centering
\begin{tabular}{lrrccccccr}
\hline
Galaxy&$T_e$& $n_e$& O/H$^1$&O/H&O/H&
O/H&O/H&N/O$^2$&O/H\\
Name&(K)&(e$^{-}cm^{-3}$)&($T_e$)&($R_{23}$)&(P)&
(N2)&(O3N2)&&(OTH)$^3$\\
\hline
iiizw12    &  9042.7 &   100.0$^4$ &  8.34 &  8.69 &  8.34 &  8.65 &  8.55 & -1.07 &8.53$^a$ \\
haro15     &  9652.4 &   100.0$^4$ &  8.33 &  8.60 &  8.33 &  8.78 &  8.48 & -0.84 &8.56$^a$\\
iiizw33    &  9539.6 &   199.7 &  8.37 &  8.58 &  8.26 &  8.53 &  8.40 & -1.27 & \\
vzw155     &  6636.2 &  1545.1 &  8.64 &  8.89 &  8.55 &  8.81 &  8.78 & -0.89 & \\
iiizw42    &     0.0 &   209.8 &  0.00 &  0.00 &  0.00 &  8.82 &  0.00 &  0.00 & \\
iiizw43    &  6412.1 &   182.5 &  8.62 &  8.90 &  8.60 &  8.78 &  8.66 & -0.82 &8.99$^b$ \\
iizw23     &  7396.1 &  3208.8 &  8.64 &  8.80 &  8.52 &  8.82 &  8.59 & -0.92 &8.55$^c$ \\
iizw28     &  9519.2 &   100.0$^4$ &  8.32 &  8.64 &  8.33 &  8.63 &  8.47 & -1.08 & \\
iizw33     & 14707.6 &   100.0$^4$ &  7.99 &  8.34 &  8.01 &  8.40 &  8.29 & -1.40 &8.03$^c$ \\
iizw40     & 13679.5 &    80.1 &  8.07 &  8.24 &  7.94 &  7.90 &  0.00 & -1.44 &8.09$^c$ \\
mrk5       & 12230.3 &   100.0$^4$ &  8.05 &  8.00 &  7.82 &  8.15 &  0.00 & -1.36 &8.04$^b$ \\
viizw153   & 10843.7 &   281.6 &  8.23 &  8.54 &  8.23 &  8.51 &  8.37 & -1.24 & \\
viizw156   &  7827.8 &   135.9 &  8.51 &  8.75 &  8.51 &  8.56 &  8.42 & -1.09 & \\
haro1      &  7057.9 &   100.0$^4$ &  8.53 &  8.85 &  8.52 &  8.57 &  8.58 & -1.14 & \\
mrk385     &  6883.9 &    21.0 &  8.55 &  8.86 &  8.54 &  8.85 &  8.70 & -0.76 & \\
mrk390     &  9674.0 &   712.9 &  8.35 &  8.61 &  8.28 &  8.58 &  8.45 & -1.21 & \\
zw0855     &  8894.7 &   504.2 &  8.43 &  8.65 &  8.41 &  8.49 &  8.35 & -1.22 & \\
mrk105     &  6526.7 &   145.4 &  8.62 &  8.89 &  8.62 &  8.71 &  8.59 & -0.88 & \\
izw18      & 21334.8 &   104.7 &  7.15 &  7.45 &  7.34 &  7.61 &  0.00 & -1.54 &7.16$^b$,7.20$^e$ \\
mrk402     & 13301.8 &   100.0$^4$ &  8.01 &  8.49 &  8.23 &  8.47 &  8.30 & -1.17 & \\
haro22     & 15790.0 &  1112.7 &  7.88 &  8.22 &  8.06 &  8.12 &  0.00 & -1.56 & \\
haro23     &  9351.0 &    44.2 &  8.35 &  8.63 &  8.36 &  8.53 &  8.39 & -1.18 &8.40$^d$ \\
iizw44     &  6128.0 &   100.0$^4$ &  8.63 &  8.93 &  8.62 &  8.88 &  8.82 & -0.68 & \\
haro2      &  9510.3 &   590.2 &  8.35 &  8.63 &  8.33 &  8.54 &  8.42 & -1.22 &8.40$^e$ \\
mrk148     &  9889.8 &    68.2 &  8.27 &  8.63 &  8.28 &  8.80 &  8.60 & -0.88 & \\
haro3      & 11645.5 &   194.8 &  8.13 &  8.55 &  8.35 &  8.33 &  8.22 & -1.29 &8.25$^b$,8.37$^d$ \\
haro25     &  8491.9 &   617.3 &  8.48 &  8.67 &  8.46 &  8.51 &  8.34 & -1.16 & \\
mrk1267    &  6709.0 &   100.0$^4$ &  8.60 &  8.87 &  8.60 &  8.85 &  8.64 & -0.69 &8.51$^a$ \\
haro4      & 14518.0 &   154.5 &  7.85 &  7.96 &  7.76 &  7.87 &  0.00 & -1.48 & 
7.81$^b$,7.80$^d$,7.80$^e$\\
mrk169     &  8663.5 &  7001.9 &  8.62 &  8.66 &  8.44 &  8.86 &  8.51 & -0.93 & \\
haro27     & 10841.3 &   182.9 &  8.20 &  8.57 &  8.21 &  8.56 &  8.46 & -1.20 & \\
mrk201     &  6330.6 &   602.3 &  8.68 &  8.90 &  8.66 &  8.98 &  8.69 & -0.50 &8.81$^a$ \\
haro28     & 12648.8 &   362.3 &  8.10 &  8.46 &  8.11 &  8.52 &  8.40 & -1.26 & \\
haro8      & 11645.5 &   138.4 &  8.12 &  8.56 &  8.38 &  8.25 &  8.18 & -1.35 & \\
\hline
\end{tabular}
\\

1: O/H = \zoh
;
2: N/O = log(N/O);
3: OTH is the oxygen abundance of BCGs from the previous studies,
derived by the $T_e$ method. References: 
a) Storchi-Bergmann \etal (1994);
b) Izotov \& Thuan (2004);
c) Vacca \& Conti (1992);
d) Hunter \& Hoffman (1999);
e) Mas-Hesse \& Kunth (1999).
4: $n_e$=100$e^-cm^{-3}$ is an assumption value.
\\
\end{table*}

\setcounter{table}{0}
\begin{table*}
\centering
\caption{\it Continued}
\begin{tabular}{lrrccccccr}
\hline
Galaxy&$T_e$& $n_e$& O/H$^1$&O/H&O/H&
O/H&O/H&N/O$^2$&O/H\\
Name&(K)&(e$^{-}cm^{-3}$)&($T_e$)&($R_{23}$)&(P)&
(N2)&(O3N2)&&(OTH)$^3$\\
\hline
haro29     & 15407.9 &   151.8 &  7.85 &  8.02 &  7.80 &  7.70 &  0.00 & -1.54 &
7.81$^b$,7.84$^d$,7.80$^e$ \\
mrk213     &  6445.0 &  1138.0 &  8.72 &  8.88 &  8.71 &  9.03 &  8.65 & -0.37 & \\
mrk215     &  6547.2 &    50.6 &  8.57 &  8.90 &  8.55 &  8.84 &  8.78 & -0.78 & \\
haro32     &  9842.9 &    26.1 &  8.29 &  8.62 &  8.30 &  8.60 &  8.45 & -1.12 & \\
haro33     & 11363.7 &  1075.8 &  8.28 &  8.43 &  8.17 &  8.29 &  8.21 & -1.57 & \\
haro34     &  6467.8 &   473.5 &  8.60 &  8.90 &  8.56 &  8.78 &  8.73 & -0.89 & \\
haro36     &  9424.0 &   100.0$^4$ &  8.37 &  8.61 &  8.37 &  8.31 &  8.27 & -1.45 &8.42$^d$ \\
haro35     &  8026.5 &    29.6 &  8.47 &  8.74 &  8.47 &  8.57 &  8.45 & -1.11 & \\
haro37     &  8075.5 &   100.0$^4$ &  8.46 &  8.75 &  8.46 &  8.64 &  8.49 & -1.04 & \\
mrk57      &  8284.5 &   100.0$^4$ &  8.41 &  8.75 &  8.41 &  8.73 &  8.59 & -0.96 & \\
mrk235     &  6624.8 &   243.9 &  8.60 &  8.88 &  8.58 &  8.73 &  8.64 & -0.90 & \\
mrk241     &  5918.3 &    98.8 &  8.65 &  8.95 &  8.64 &  8.79 &  8.73 & -0.80 & \\
izw53      &  6831.4 &   100.0$^4$ &  8.59 &  8.86 &  8.58 &  8.74 &  8.60 & -0.86 & \\
izw56      &  8383.4 &   392.0 &  8.41 &  8.74 &  8.39 &  8.94 &  8.71 & -0.57 & \\
haro38     & 13679.5 &   100.0$^4$ &  8.02 &  8.23 &  8.10 &  8.18 &  8.14 & -1.53 & \\
mrk275     &  8797.0 &  2400.9 &  8.49 &  8.67 &  8.41 &  8.56 &  8.40 & -1.24 & \\
haro39     & 10317.9 &   950.7 &  8.30 &  8.57 &  8.28 &  8.25 &  8.26 & -1.62 & \\
haro42     & 13442.2 &   135.9 &  7.98 &  8.52 &  8.33 &  8.38 &  8.22 & -1.15 & \\
haro43     &  8744.2 &  4734.7 &  8.56 &  8.67 &  8.42 &  8.46 &  8.36 & -1.52 & \\
haro44     & 13489.4 &   100.0$^4$ &  8.02 &  8.21 &  8.07 &  8.18 &  8.13 & -1.51 & \\
iizw70     & 15087.8 &   160.2 &  7.69 &  7.80 &  7.69 &  8.07 &  8.14 & -1.38 &8.00$^e$ \\
iizw71     & 10900.2 &   203.7 &  8.24 &  8.52 &  8.24 &  8.64 &  8.41 & -1.06 & \\
izw97      &  7478.8 &   100.0$^4$ &  8.51 &  8.80 &  8.51 &  8.66 &  8.54 & -1.01 & \\
izw101     &     0.0 &   100.0$^4$ &  0.00 &  0.00 &  0.00 &  8.61 &  0.00 &  0.00 & \\
izw117     &  6902.8 &   193.9 &  8.55 &  8.87 &  8.53 &  8.74 &  8.67 & -0.93 & \\
izw123     & 15035.1 &   144.8 &  7.86 &  8.07 &  7.86 &  8.16 &  0.00 & -1.29 &8.46$^a$,8.10$^b$ \\
mrk297     &  9007.4 &    75.1 &  8.34 &  8.69 &  8.35 &  8.69 &  8.56 & -1.01 & \\
izw159     & 12621.6 &  3267.1 &  8.12 &  8.54 &  8.31 &  8.38 &  8.26 & -1.37 & \\
izw166     &  9175.6 &    77.1 &  8.37 &  8.64 &  8.37 &  8.57 &  8.41 & -1.12 &8.47$^a$ \\
mrk893     & 11263.6 &   468.2 &  8.19 &  8.54 &  8.19 &  8.64 &  8.47 & -1.10 & \\
izw191     &  6306.9 &   100.0$^4$ &  8.60 &  8.92 &  8.58 &  8.81 &  8.77 & -0.82 & \\
ivzw93     &  9993.0 &  1279.2 &  8.35 &  8.58 &  8.31 &  8.31 &  8.27 & -1.54 & \\
mrk303     &  7612.3 &   100.0$^4$ &  8.48 &  8.80 &  8.48 &  8.85 &  8.76 & -0.77 & \\
zw2220     &  7116.7 &   224.9 &  8.55 &  8.83 &  8.54 &  8.80 &  8.62 & -0.80 & \\
mrk314     & 10946.7 &   100.0$^4$ &  8.26 &  8.49 &  8.25 &  8.35 &  8.24 & -1.39 & \\
ivzw142    &  7041.9 &   100.0$^4$ &  8.55 &  8.85 &  8.54 &  8.85 &  8.64 & -0.75 & \\
ivzw149    &  9855.7 &   491.9 &  8.32 &  8.60 &  8.31 &  8.64 &  8.44 & -1.07 &8.48$^a$ \\
zw2335     &  8164.4 &  6515.8 &  8.66 &  8.75 &  8.44 &  8.67 &  8.53 & -1.26 &8.65$^d$ \\
\hline
\end{tabular}
\end{table*}